\documentclass{ws-ijmpa}

\begin{document}

\markboth{V.~I.~Kuksa, R.~S.~Pasechnik} {Near-threshold W-pair
production in the model of unstable particles with smeared mass}

%
\catchline{}{}{}{}{}
%

\title{NEAR-THRESHOLD W-PAIR PRODUCTION IN THE MODEL \\
OF UNSTABLE PARTICLES WITH SMEARED MASS}

\author{V.~I.~KUKSA\footnote{kuksa@list.ru}}

\address{Institute of Physics, Southern Federal University, Rostov-on-Don 344090, Russia}

\author{R.~S.~PASECHNIK\footnote{rpasech@theor.jinr.ru}}

\address{Bogoliubov Laboratory of Theoretical Physics, JINR,
Dubna 141980, Russia}

\maketitle

\begin{history}
\received{Day Month Year} \revised{Day Month Year}
\end{history}

\begin{abstract}
Near-threshold production of charged boson pairs is considered
within the framework of the model of unstable particles with smeared
mass. The results of calculations are in good agreement with LEP II
data and Monte-Carlo simulations. Suggested approach significantly
simplifies calculations with respect to standard perturbative one.

\end{abstract}

\ccode{PACS number: 11.10.St}

\section{Introduction}

The measurements of $W$-pair production at LEP II provided us with
an important information about the mass of $W$ boson and non-abelian
triple gauge-boson couplings. To extract the exact information from
$W$-pair production we have to calculate the radiative corrections
(RC's), which give a noticeable contribution to the cross-section.
Ideally, one would like to have the full RC's to the process
$e^+e^-\to W^+W^-\to 4f$. In practice, this problem is very
complicated and can not be considered analytically. For discussion
of the LEP II situation and strategy it is useful to distinguish
three levels of sophistication in the description of the $W$-pair
production [\refcite{1,2}]:

1) On-shell $W$-pair production, $e^+e^-\to W^+W^-$, with
consequent on-shell $W$ decays. All $O(\alpha)$ RC's to these
processes are known.

2) Off-shell production of $W$ pairs, which then decay into
four fermions. Full set of RC's is very bulky for the
analytical observation and analysis.

3) Full process $e^+e^-\to 4f$ with an account of the complete set
of the $O(\alpha)$ corrections. This problem leads to the additional
diagrams with the same final states, and complete electroweak $O(\alpha)$
corrections are described by many thousands diagrams.

On-shell $W$-pair production was considered in
Refs.~[\refcite{1,2,3}], where the cross-section of the process
$e^+e^-\to W^+W^-$ was given. At tree level, this process is
described by three diagrams shown in Fig.~1. The complete
$O(\alpha)$ radiative corrections, comprising the virtual one-loop
corrections and real-photon bremsstrahlung, were calculated and
represented in Refs.~[\refcite{4}]-[\refcite{11}]. The description
of the on-shell $W$-pair production and consequent decays with an
account of RC's was fulfilled in
Refs.~[\refcite{12}]-[\refcite{18}]. Off-shell production of
$W$-pairs, which then decay into four fermions, was considered in
Ref.~[\refcite{19}].

In description of the $W$- and $Z$-pairs production we should take
into consideration the fact that the gauge bosons are not stable
particles and the real process is not $e^+e^-\to W^+W^-,ZZ$
[\refcite{2}]. This is only an approximation with a level of
goodness, which may depend on several factors, while the real
process is $e^+e^-\to W^+W^-,ZZ\to 4f$. There are many papers,
devoted to comprehensive analysis and description of all possible
processes with the four-fermion final states. Because of a large
number of diagrams, describing these processes, the classification
scheme was applied in Refs.~[\refcite{20}]-[\refcite{23}]. The
possible processes are divided into three classes: charge current
(CC), neutral current (NC) and mixed current (MIX). Born processes
$e^+e^-\to W^+W^-,ZZ$ are designated as CC03 and NC02, which
correspond to three charge current and two neutral current diagrams.
According to this classification the off-shell $W$-pair production
with consequent $W$ decay can be described in the framework of the
Double-Pole Approximation (DPA) [\refcite{23}]-[\refcite{26}]. The
DPA selects only diagrams with two nearly resonant $W$ bosons and
the number of graphs is considerably reduced [\refcite{23}].

Complete description of the total set of $4f$-production processes
with an account of RC's is not analytically available due to a huge
number of diagrams and presence of non-factorable corrections. But
the complete EW $O(\alpha)$ corrections have been calculated for
some exclusive processes, for instance, for the processes
$e^+e^-\to\nu_{\tau}\tau^+\mu^-\bar{\nu}_{\mu},\,u\bar{d}\mu^-\bar{\nu}_{\mu}$,
and $u\bar{d}s\bar{c}$ [\refcite{26a,26b}].  Because of complexity
of the problem, some approximation schemes are practically applied,
namely, Semi-Analytical Approximation (SAA) [\refcite{2,26ab}],
improved Born approximation [\refcite{26bb}], an asymptotic
expansion in powers of the coupling constant of the cross-section
[\refcite{26bbb}], fermion-loop scheme, etc. (see Introduction in
Ref.~[\refcite{26a,26b}]). There are many computer tools of
calculations, for instance, Monte-Carlo (MC) simulations, such as
RacoonWW [\refcite{26b,26c,26d}] and YFSWW [\refcite{26e,26f,28}].
All above mentioned methods are based on the traditional quantum
field theory of unstable particles (UP's) [\refcite{2}]. At the same
time, there are some alternative approaches for description of the
UP's, such as the effective theory of UP's
[\refcite{29}]-[\refcite{31}] and the model of UP with smeared mass
[\refcite{32,33}].

In this paper, we suggest the description of the near-threshold
$W$-pairs production within the framework of the model of UP with
smeared mass initially proposed in Ref.~[\refcite{32}]. The model is
based on the time-energy uncertainty relation $\Delta E\cdot \Delta
t\sim 1\,(c=\hbar=1)$. It follows from the equation of motion in the
Heisenberg representation which describes the evolution of the
non-stationary quantum system [\refcite{33a}]. In the case of the
unstable particles, $\Delta t$ is the lifetime and $\Delta E$ is the
value of the mass smearing $\Delta m$ in the rest-frame system
[\refcite{32,33a,33b}]. In the model under consideration the UP is
described by a state with smeared (fuzzed) mass in accordance with
the uncertainty relation. So, the processes $e^+e^-\to W^+W^-,\, ZZ$
are described in a traditional way, i. e. in a stable particle
approximation, but the phase space is calculated for the states on
the smeared mass-shell. In the framework of the model, full process
$e^+e^-\to W^+W^-,\, ZZ \to 4f$ is divided into two stages
$e^+e^-\to W^+W^-,\, ZZ$ and $W^+W^-,\, ZZ \to 4f$ due to exact
factorization at tree level [\refcite{34}]-[\refcite{35a}]. For
description of the bosons in the final state we use the model
polarization matrix which differs from the standard one
[\refcite{33,35a}] (see the next section).

The model was applied for description of the Finite-Width Effects
(FWE's) in many low- and high-energy processes involving the UP with
large width [\refcite{33}], [\refcite{34}]-[\refcite{35b}]. In
particular, the approach was successfully applied to the process
$e^+e^-\to ZZ$ in Ref.~[\refcite{33}]. In this paper it was shown
that the results of the model are in good agreement with the LEP II
data and turned out to be very close to the corresponding results of
MC simulations. So, it is reasonable to apply the same method for
the case of $W$-pair production.

It should be also noted that the model under consideration directly
leads to calculation schemes which are in close analogy with such
standard approaches as the Convolution Method (CM), Narrow-Width
Approximation (NWA) and SAA. However, the model treatment has some
noticeable distinctions which are discussed in detail in
Refs.~[\refcite{34}]-[\refcite{35a}] (see also the next section).
The principal distinction between the standard and model treatment
of the FWE's takes place in description of UP with large width
[\refcite{35a}] and the mass splitting in the neutral meson systems
[\refcite{35b}].

The paper is organized as follows. In the second section, we give a
short description of the model and define the status of our
calculations. The cross-section of the process $e^+e^-\to W^+W^-$ at
tree level is derived in the framework of the model [\refcite{32}]
in this section. Section 3 contains the calculation strategy with
taking into account of the radiative corrections. In this section we
also represent the results of our calculations, MC simulations and
LEP II data. Some conclusions concerning the applicability of the
method were made in the last section. We note that the aim of this
investigation is to test the model approach for the case of
near-threshold boson-pair production and analyze the possibility of
its improvement by an accounting of the radiative corrections which
are not included into the effective field of UP.

\section{The model cross-section of the near-threshold $W$-pair production at tree level}

Firstly, we give a short description of the model of UP with smeared
mass. The model wave function of the UP is
\begin{equation}\label{2.a}
 \Phi_a(x)=\int\Phi_a(x,\mu)\omega(\mu)d\mu,
\end{equation}
where $\Phi_a(x,\mu)$ is the standard spectral component, which
defines a particle with a fixed mass squared $m^2=\mu$ in the Stable
Particle Approximation (SPA). The weight function $\omega(\mu)$ is
formed by the self-energy interactions of UP with vacuum
fluctuations and decay products. This function describes the smeared
(fuzzed) mass-shell of UP. Thus, the smearing of mass is caused, on
the one hand, by instability according to formal uncertainty
relation and, on the other hand, by stochastic interaction of UP
with the electro-week vacuum fluctuations [\refcite{35}].

The commutative relations for the model operators have an additional
$\delta$-function
\begin{equation}\label{2.b}
 [\dot{\Phi}^{-}_{\alpha}(\bar{k},\mu),\,\Phi^{+}_{\beta}(\bar{q},\mu')]_{\pm}
 =\delta(\mu-\mu') \delta(\bar{k}-\bar{q})\delta_{\alpha\beta},
\end{equation}
where subscripts ``$\pm$'' correspond to fermion and boson fields,
respectively. The presence of $\delta(\mu-\mu')$ in Eq.~(\ref{2.b})
means the following assumption: the acts of creation and
annihilation of the particles with various $\mu$ (the random mass
squared) do not interfere. Thus, the parameter $\mu$ has the status
of physically distinguishable value of a random $m^2$.

 The model Green functions for the vector and spinor fields in momentum
representation have the convolution form:
\begin{equation}\label{2.c}
 D_{mn}(k)=-i\int \frac{g_{mn}-k_mk_n/\mu}{k^2-\mu+i\epsilon} \rho(\mu)d\mu\,,
\end{equation}
and
\begin{equation}\label{2.d}
 \hat{D}(k)=i\int \frac{\hat{k}+k}{k^2-\mu+i\epsilon} \rho(\mu)d\mu\,,
\end{equation}
where $\rho(\mu)=|\omega(\mu)|^2$.

Further, we consider the model amplitude for the simplest processes
with UP in the initial or final state. The expression for scalar
field is
\begin{equation}\label{2.e}
 \phi^{\pm}(x)=\frac{1}{(2\pi)^{3/2}}\int\omega(\mu)d\mu\int\frac{a^{\pm}(\bar{q},\mu)}
 {\sqrt{2q^0_{\mu}}}e^{\pm iqx}d\bar{q}\,,
\end{equation}
where $q^0_{\mu}=\sqrt{\bar{q}^2+\mu}$ and $a^{\pm}(\bar{q},\mu)$
are the creation or annihilation operators of UP with the momentum
$q$ and mass squared $m^2=\mu$. Taking into account Eq.~(\ref{2.b}),
one can get
\begin{equation}\label{2.f}
 [\dot{a}^{-}(\bar{k},\mu),\phi^{+}(x)]_{-}\,,\; [\phi^{-}(x), \dot{a}^{+}(\bar{k},\mu)]_{-}
 =\frac{\omega(\mu)}{(2\pi)^{3/2}\sqrt{2k^0_{\mu}}}e^{\pm ikx}\,,
\end{equation}
where $k^0_{\mu}=\sqrt{\bar{k}^2+\mu}$. The expressions (\ref{2.f})
differ from the standard ones by the factor $\omega(\mu)$ only. From
this result it follows that, if $\dot{a}^{+}(k,\mu)|0\rangle$ and
$\langle0|\dot{a}^{-}(k,\mu)$ define UP with the mass $m=\sqrt{\mu}$
and momentum $k$ in the initial or final states, then the amplitude
for the transition $\Phi\rightarrow\phi\phi_1$ is
\begin{equation}\label{2.g}
 A(k,\mu)=\omega(\mu)A^{st}(k,\mu)\,,
\end{equation}
where $A^{st}(k,\mu)$ is the amplitude in the SPA. This amplitude is
calculated in the standard way and can include the higher
corrections. Moreover, it can be an effective amplitude for the
processes with hadron participation. From Eq.~(\ref{2.g}) it follows
that the differential (over $\mu$) probability of transition is
$dP(k,\mu)=\rho(\mu)|A(k,\mu)|^2d\mu$.

To define the transition probability of the process
$\Phi\rightarrow\phi\phi_1$, where $\phi$ is UP with a large width,
we should take into account the status of parameter $\mu$ as a
physically distinguishable value, which follows from
Eq.~(\ref{2.b}). Thus, the differential (over $k$) probability is
\begin{equation}\label{2.h}
 d\Gamma(k)=\int d\Gamma^{st}(k,\mu)\rho(\mu)d\mu\,.
\end{equation}
In Eq.~(\ref{2.h}) the differential probability
$d\Gamma^{st}(k,\mu)$ is defined in the standard way (the SPA).

If there are two UP's with large widths in the final state of decay
$\Phi\rightarrow\phi_1\phi_2$, then in analogy with the previous
case one can get the double convolution formula:
\begin{equation}\label{2.x}
 \Gamma(m_{\Phi})=\int\int\Gamma^{st}(m_{\Phi};\mu_1,\mu_2)\rho_1(\mu_1)\rho_2(\mu_2)d\mu_1
 d\mu_2\,.
\end{equation}
The polarization matrix for the case of vector UP in the final state
has the form
\begin{equation}\label{2.y}
 \sum_{e} e_m(q)e^{*}_n(q)=-g_{mn}+q_mq_n/\mu\,,
\end{equation}
In the case of spinor UP in the final state we have
\begin{equation}\label{2.z}
 \sum_{\nu} u^{\nu,\pm}_{\alpha}(q)\bar{u}^{\nu,\mp}_{\beta}(q)=\frac{1}{2q^0_{\mu}}
 (\hat{q}\mp\sqrt{\mu})_{\alpha\beta}\,,
\end{equation}
where the summation over polarizations is implied and
$q^0_{\mu}=\sqrt{\bar{q}^2+\mu}$. The same relations take place for
the initial states, however, one have to average over their
polarizations.

The most important element of the model is the probability density
$\rho(\mu)$ which describes the smearing of UP mass. The various
definitions of $\rho(\mu)$ were discussed in Ref.~[\refcite{35}],
where the Lorentzian (Breit-Wigner type), Gaussian and
phenomenological distributions have been considered. The Lorentzian
distribution was derived by matching the model propagators
(\ref{2.c}), (\ref{2.d}) and standard dressed ones in the
Breit-Wigner form.

One of the important properties of the model is the exact
factorization of the processes with the UP in an intermediate state.
In the frame of the effective theory of UP, which follows from the
model, the factorization leads to the convolution formula for the
decay rate [\refcite{32a}] and factorized formula for the
cross-section [\refcite{34}]. These results are derived by
straightforward calculations at tree level without any
approximations. So, the model provides the formal basis for the CM,
SAA and NWA, which are the approximate approaches in the traditional
treatment. The generalization of the factorization method to the
complicated processes of scattering and decays with two or more UP
in intermediate states was considered in Ref.~[\refcite{35a}]. This
method can be applied for description of the boson-pair production
and decays in the factorized form for double-pole set of diagrams
(see comment to tree-level result at the end of this section).

In this work, we use two principal elements of the model
[\refcite{32,35}] -- the convolution structure of the transition
probability (an analog of the expression (\ref{2.x})) and the
polarization matrix for UP in the final state (\ref{2.y}). Using
these expressions, we get the model Born cross-section of $W$-pair
production in the following form
\begin{equation}\label{2.3}
\sigma^{B}_{WW}(s)=\int\int\sigma^{B}_{WW}(s;\mu_1,\mu_2)\rho_1(\mu_1)\rho_2(\mu_2)\,d\mu_1\,d\mu_2\,,
\end{equation}
where $\sigma^{B}_{WW}(s;\mu_1,\mu_2)$ is the Born cross-section
which is calculated in the standard way for fixed bosons masses
$\mu_1=m^2_1$ and $\mu_2=m^2_2$ (SPA).
\begin{figure}[h!]
\centerline{\epsfig{file=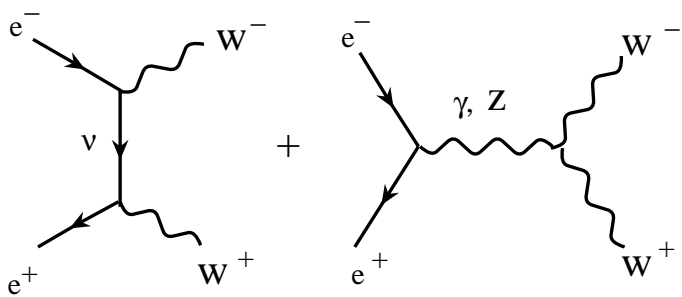,width=9cm}} \caption{Feynman
diagrams for the process $e^+e^-\to W^+W^-$.} \label{fig:FD}
\end{figure}

The Born cross-section is defined by the sum of two diagrams shown
in Fig.~1 and can be represented as
\begin{equation}\label{2.4}
\sigma^{B}_{WW}(s;x_1,x_2)=\frac{\pi\alpha^2}{128 s \sin^4\theta_{W}}F(s;x_1,x_2),
\end{equation}
where dimensionless function $F(s;x_1,x_2)$ is defined by the
expression
\begin{align}\label{2.5}
F(s;x_1,x_2)&=\frac{16}{3(a^2-b^2)(1-x_Z)^2}\{3(a^2-b^2)(a^2-b^2+2(1+a))(1-x_Z)^2 L(a,b)\notag\\
            &+x_Z\cos(2\theta_W)[3(b^4-2ab^2(2+a)+a^3(4+a))(1-x_Z)L(a,b)\notag\\
            &+2\lambda(a,b)(2b^2-3a^2-10a-1)(b^2(1-2x_Z)-a(1-3x_Z)-x_Z)]\notag\\
            &+\lambda(a,b)[x^2_Z\lambda^2(a,b)\cos(4\theta_W)(2b^2-3a^2-10a-1)+12a^3z^2_Z\notag\\
            &-a^2(3b^2(3x^2_Z-2x_Z+1)-49x^2_Z+30x_Z-15)-2a(b^2(19x^2_Z-10x_Z+5)\notag\\
            &+8x^2_Z)+2b^4(3x^2_Z-2x_Z+1)-2b^2(7x^2_Z-16x_Z+8)-2x^2_Z]\}.
\end{align}
In Eq.~(\ref{2.5}) the dimensionless variables $a,b,x_1,x_2,x_Z$ and
the functions $L(a,b)$ and $\lambda(a,b)$ are defined as follows
\begin{align}\label{2.6}
&L(a,b)=\ln\biggl[\frac{1-a-\lambda(a,b)}{1-a+\lambda(a,b)}\biggr],\,\,\,\lambda(a,b)=\sqrt{1-2a+b^2},\notag\\
&x_{1,2}=\frac{\mu_{1,2}}{s},\,\,\,a=x_1+x_2,\,\,\,b=x_1-x_2,\,\,\,x_Z=\frac{M^2_Z}{s}.
\end{align}
With the help of the expressions (\ref{2.4})-(\ref{2.6}) the model Born
cross-section is represented in the following convolution form
\begin{equation}\label{2.7}
\sigma^{B}_{WW}(s)=\frac{\pi\alpha^2}{128 s \sin^4\theta_W}
\int_{0}^{1}\,dx_1\rho(x_1,s)\int_{0}^{(1-\sqrt{x_1})^2}\rho(x_2,s)F(s;x_1,x_2)\,dx_2,
\end{equation}
where in analogy with the case of $Z$-pair production [\refcite{33}]
we use the redefined dimensionless probability density of Lorentzian
type
\begin{equation}\label{2.8}
\rho(x,s)=\frac{1}{\pi}\frac{G(x,s)}{(x-x_W)^2+G^2(x,s)},\,\,\,
G(x,s)=\frac{\sqrt{\mu}\Gamma^{tot}_W}{s}=\frac{3\alpha
x}{4\sin^2\theta_W},
\end{equation}
where $x=\mu/s$, $x_W=M^2_W/s$. The expression (\ref{2.7}) turns
into the standard expression for the on-shell cross-section
$\sigma^{B}_{WW}(s)$, when $\rho(\mu)\to \delta(\mu-M^2_W)$, i.e. in
the limit of fixed $W$ mass.
\begin{figure}[h!]
\centerline{\epsfig{file=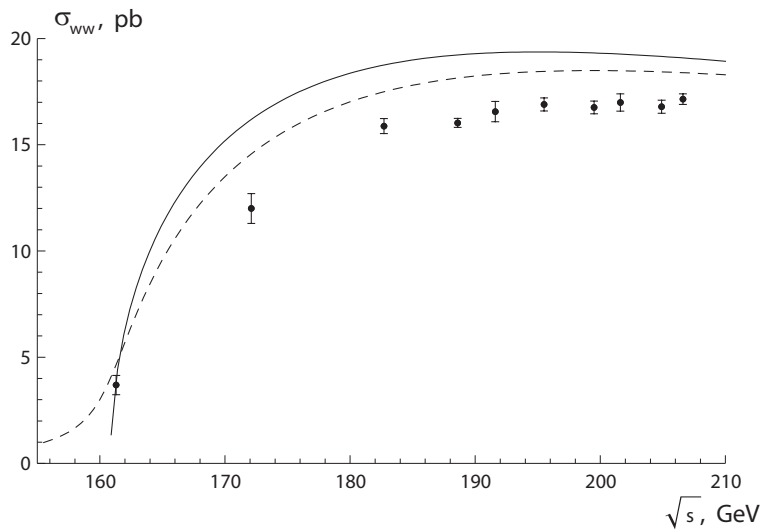,width=10cm}}
\caption{On-shell (solid line) and model (dashed line) Born
cross-section of the process $e^+e^-\to W^+W^-$.} \label{fig:FD}
\end{figure}

In Fig.~2 we represent the Born cross-section in the fixed-mass
approach (solid line) and in the smeared-mass approach (dashed
line). Besides, in this figure we show the experimental LEP2 data in
order to illustrate the necessity of radiative corrections. One can
see that the model approach leads to the smearing of the threshold,
that is to the result which is similar to the standard one with an
account of the Finite-Width Effects (FWE). In other words, the model
description of boson-pair production is in close analogy with the
standard description of the off-shell boson-pair production.
Moreover, the model convolution representation of the cross-section
in the form (\ref{2.3}) or (\ref{2.7}) is formally similar to the
SAA [\refcite{2,21}]. However, the SAA is constructed as an
approximation for the exclusive process like $e^+e^-\to W^+W^-\to
f_1\bar{f}_2f_3\bar{f}_4$, which then is generalized to the
inclusive process with full set of the $4f$ final states. This
approach is based on the approximate factorization of the total
cross-section $\sigma(e^+e^-\to f_1\bar{f}_2f_3\bar{f}_4)
\longrightarrow \sigma(e^+e^-\to W^+W^-)\mathrm{Br}(W^+\to
f_1\bar{f}_2)\mathrm{Br}(W^-\to f_3\bar{f}_4)$. In the frame of the
model [\refcite{32,35}], the factorization is exact (see also
Refs.~[\refcite{34,35a}]), and the expression (\ref{2.3}) can be
directly derived [\refcite{35a}] for the inclusive process
$e^+e^-\to \sum_{f}4f$ in the double-pole approach without any
approximations.

From Fig.~2 it follows that the use of the effective model fields,
which describe the UP with an account of the self-energy type
corrections, is not sufficient. We have to take into account the
rest radiative corrections for the realistic description of the
measured cross-section.

Now we estimate the uncertainties of the model calculations at the
effective tree level, which are caused mainly by the definition of
the function $\rho(\mu)$. As was shown in the framework of the
effective theory of UP [\refcite{34}]-[\refcite{35a}], this function
results from the factorization of full process of production and
decay of UP. Let us define the uncertainty as a deviation of the
model calculation from the standard one. The part of the model
amplitude which describes the decays $W\to l\nu_l$ is
\begin{equation}\label{2.9}
\mathit{M}^{mod}\sim\frac{(-g^{\mu\nu}+\frac{q_1^{\mu}q_1^{\nu}}{q^2_1})}{P(q^2_1)}\,\frac{(-g^{\mu'\nu'}
+\frac{q_2^{\mu'}q_2^
{\nu'}}{q^2_2})}{P(q^2_2)}\,\bar{l}_1\gamma_{\nu}(1-\gamma_5)\nu_1\cdot
\bar{\nu}_2\gamma_{\nu'}(1-\gamma_5)l_2.
\end{equation}
The standard expression for the amplitude follows from
Eq.~(\ref{2.9}) after the change $q^2_a \to M^2_W$ in the numerator
of the dressed propagators of unstable bosons $W$. We assume that
the denominators $P(q^2_a)$ in both cases are the same (in the
Breit-Wigner or complex pole form). As was mentioned above, the use
of the standard propagators does not lead to the exact factorization
even at the tree level. In the standard approach this effect takes
place in the Narrow-Width Approximation (NWA), while in the
framework of the model under consideration the factorization is
exact due to specific form of propagator's numerator
[\refcite{35a}]. Using the equality
\begin{equation}\label{2.10}
\bar{l}\hat{q}_a(1-\gamma_5)\nu_a=m_a \bar{l}(1-\gamma_5)\nu_a,
\end{equation}
where $a=1,2$, we get
\begin{equation}\label{2.11}
|\mathit{M}^{mod}|^2\sim 1+2\frac{m_1}{q_1}+2\frac{m_2}{q_2},
\end{equation}
Where $q=\sqrt{(q\cdot q)}$. The same expression takes place for the
standard amplitude squared $|\mathit{M}^{st}|^2$ after the change
$q_a\to M_W$. As a result, we have the relative deviation of the
model partial cross-section from the standard one ($m_1=m_2=m_f$):
\begin{equation}\label{2.12}
\epsilon_f \sim 4\frac{m_f}{M_W}[1-M_W\int_{m^2_f}^{s}\frac{\rho(q^2)}{q}\,dq^2].
\end{equation}

From (\ref{2.12}) with the help of the Breit-Wigner approximation
for the function $\rho(q^2)$ we find that the maximal deviation is
for heavy fermions, for instance, for the $\tau$-lepton pair and
$b,c$-quark pairs. However, in the last case the $\epsilon$ is
suppressed by small CKM elements $|U_{cb}|^2$. Thus, from
(\ref{2.12}) it follows that $\epsilon_{max}=\epsilon_{\tau}\sim
10^{-3}$, that is the uncertainty of our approach is an order of
$0.1\%$. From this simple analysis we make a conclusion that an
error, which is caused by the model approach at tree level,
noticeably less than $1\%$. So, the main uncertainty can be caused
only by the implantation of the radiative corrections into our
scheme of calculation (see the next section).

\section{The model cross-section of $W$-pair production
with radiative corrections}

In this section we discuss the strategy of the RC's accounting and
represent the final results of calculations. As it was shown in
Refs.~[\refcite{34,35}], the model description of UP is equivalent
to some effective theory of UP, which includes the self-energy type
RC's in all orders of perturbation theory. Moreover, the UP is the
non-perturbative object in the vicinity of the resonance. So, the
traditional program of RC's calculation is not valid in the
framework of the model. We have no well defined set of the diagrams
which is gauge invariant and renormalized. The model of UP
[\refcite{32,35}] is effective and not gauge one, and we have no any
rigid criteria for definition of such a set. So, we keep the
strategy which is based on the simple phenomenology and was
successfully applied in the case of $Z$-pair production
[\refcite{33}].

We do not take into account any corrections to the final states $W$,
because of the effective nature of these states in the framework of
the model. We use the effective coupling $\alpha (M_W)=1/127.9$ in
the vertex with the final $W$-states and $\alpha=1/137$ in the RC's.
So, the principal part of the vertex corrections is effectively
included into the coupling, and the low-energy behavior of the
bremsstrahlung and radiative corrections to the initial states is
taken into consideration.

The set of corrections, caused by the final state interactions in
the two $s$-channel diagrams in Fig.~1, is included into the
effective coupling $\alpha(M_W)$. The principal part of the
so-called Coulomb singularity contributions, which were considered
in Refs.~[\refcite{1}], [\refcite{26a}] and [\refcite{26bbb}], can
be also absorbed by the effective coupling. The one-loop calculation
shows that this correction gives from $5.7\%$ at the threshold to
1.8\% at 190 GeV [\refcite{1}], while the total change of the
effective coupling $\alpha(M_W)$ with respect to $\alpha$ is near
$7\%$. In the calculation we explicitly take into account the
$O(\alpha)$ corrections including soft and hard bremsstrahlung,
which are not described by the model and by the effective coupling.
The real and virtual electromagnetic radiation should enter into the
set of these RC's and mutually compensates the total IR divergences.

The program of RC's calculations, which is similar to above
discussed one, was fulfilled in the series of papers (see, for
example, Ref.~[\refcite{11}] and references therein) for the case of
the on-shell $W$-pair production (the limit of fixed masses
$\mu_1=\mu_2=M^2_W$). The analytical expression for these
corrections is represented in compact and convenient form in
Ref.~[\refcite{11}]. We generalized this expression to the case of
smeared-shell $W$-pair production, that is for arbitrary values of
mass parameters $\mu_k$, and applied it in our calculations. As a
result, we get the cross-section $\sigma_{WW}(s;\mu_1,\mu_2)$ for
the case of $W(\mu_1)$ and $W(\mu_2)$ production including above
described corrections in the following form (see also
Ref.~[\refcite{11}])
\begin{equation}\label{3.1}
\sigma_{WW}(s;\mu_1,\mu_2)= \int_{0}^{k_{max}}
\rho_{\gamma}(k)\sigma^{B}_{WW}(s(1-k),\mu_1,\mu_2)\,dk\,,
\end{equation}
where $\rho_{\gamma}(k)$ is the photon radiation spectrum
[\refcite{36}]-[\refcite{38}], $k=E_{\gamma}/E_b$ is the photon
energy in units of beam energy and $s(1-k)$ is the effective $s$
available for the $W$-pair production after the photon has been
emitted [\refcite{11}]. In the case of the on-shell $W$-pair
production ($\mu_1=\mu_2=M^2_W$) the value $k_{max}=1-4M^2_W/s$ is
the maximal part of photon energy. The generalization of this value
to the case $\mu_1\ne\mu_2$ leads to
\begin{equation}\label{3.2}
k_{max}=1-2\frac{\mu_1+\mu_2}{s}+\frac{(\mu_1-\mu_2)^2}{s^2}\equiv \lambda^2(\mu_1,\mu_2;s).
\end{equation}
The photon distribution function is written in the form
[\refcite{11}]
\begin{equation}\label{3.3}
\rho_{\gamma}(k)=\beta k^{\beta-1}(1+\delta^{v+s}_1+...)+\delta^h_1+...,
\end{equation}
where we keep $O(\alpha)$ corrections only (i.e. $\delta_{n>1}=0$).
The corresponding corrections are given by ($v+s=$ virtual+soft,
$h=$ hard) [\refcite{11}]:
\begin{align}\label{3.4}
&\beta=\frac{2\alpha}{\pi}(L-1),\,\,\,L=\ln\frac{s}{m^2_e},\,\,\,
\alpha=\frac{1}{137};\notag\\
&\delta^{v+s}_1=\frac{\alpha}{\pi}(\frac{3}{2}L+\frac{\pi^2}{3}-2),\,\,\,
\delta^h_1=\frac{\alpha}{\pi}(1-L)(2-k).
\end{align}

Finally, we get the corrected expression for the cross-section of
$W$-pair production in the form
\begin{align}\label{3.5}
\sigma_{WW}(s)=&\frac{\pi\alpha^2(M_W)k_{QCD}}{128 s
\sin^4\theta_W(M_W)}\int_{0}^{1}dx_1\int_0^{(1-\sqrt{x_1})^2}dx_2
\int_{0}^{\lambda^2(x_1,x_2)}\frac{dk}{1-k}\rho_{\gamma}(k)\notag\\
&\rho(x_1,s(1-k))\rho(x_2,s(1-k))F(s(1-k);x_1,x_2).
\end{align}
where the functions $F,\,\rho,\,\rho_{\gamma}$ were defined before in
Eqs.~(\ref{2.5}), (\ref{2.8}), (\ref{3.3}) and we also take into account the
effective QCD correction factor $k_{QCD}=1+0.133/\pi$
[\refcite{39}].
\begin{figure}[h!]
\centerline{\epsfig{file=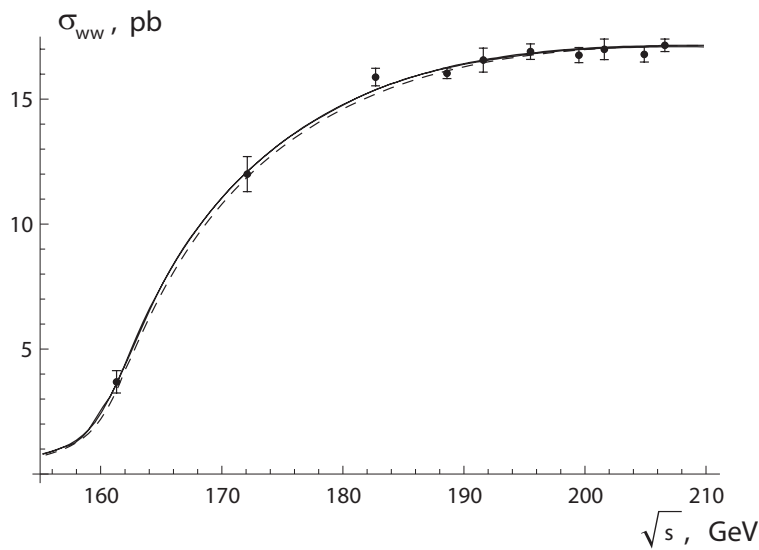,width=10cm}}  \caption{Model
(dashed line) and Monte-Carlo RacconWW and YFSWW (solid lines)
cross-sections of the process $e^+e^-\to W^+W^-$.} \label{fig:FD}
\end{figure}

The model cross-section $\sigma_{WW}(s)$ was calculated numerically
and represented in Fig.~3 as a function of $s$ by dashed line. The
results of MC simulations, RacconWW [\refcite{26c,26d}] and YFSWW
[\refcite{26e,26f}], are represented for comparison by two barely
distinguishable solid lines, and the experimental LEP II data
[\refcite{40}] are given with the corresponding error bars. From
Fig.~3, one can see that the model cross-section with RC's is in
good agreement with the experimental data. Moreover, the deviation
of the model from MC curves is significantly less then the
experimental errors ($\lesssim1\%$).

In the previous section, we have got an estimation of the tree level
uncertainty which turned out to be significantly less than the value
of radiative corrections. So, the total uncertainty of the model
approach mainly depends on the set of the RC's which we take into
account. From the above described strategy of the RC's accounting,
it follows that the principal value of the error can be caused by
the part of the Coulomb corrections, which follows from the box
diagram [\refcite{1}] and gives most likely less then 1\%, and by
the non-factorable corrections, which destroy the convolution
structure of the total cross-section. A comparison between the DPA
and the predictions based on the full $O(\alpha)$ corrections
reveals differences in the relative corrections $\lesssim 0.5\%$
[\refcite{26a}] and 0.9\% [\refcite{41}].

From the results of our calculations it follows, that the model
approach provides the accuracy which is sufficient for the LEP II
data description in the near-threshold energy range. Besides, the
contribution of the non-factorable corrections in the cross section
is less than the experimental errors. However, the value of all
non-considered corrections can be maximally an order of LEP II
uncertainty of the total cross-section, and this point needs an
additional consideration. Moreover, the actual status of the
calculations concerns rather the testing of our approach than the
tool for precise investigations. But, we believe that the approach
due its simplicity and physical transparency can provide the basis
for construction of such a tool.

\section{Conclusions}

A large number of the first order Feynman diagrams, which contribute
to the production of four fermions in $e^+e^-$ interactions, depends
on the specific final states. So, the detailed classification was
suggested for the description of these processes. The most important
request for $WW$ physics concerns the $O(\alpha)$ radiative
corrections in the DPA. Inclusion of the complete EW corrections
significantly complicates the calculations which became not
available in the analytical form in the case of the full set of
$4f$-processes. So, the various approximation schemes have been
worked out together with development of the MC simulations.

In this paper, we applied the model of UP with smeared mass for
description of the $W$-pair production. The model describes the
process $e^+e^-\to W^+W^-$ as $W$-pair production, where $W's$ are
on the smeared mass-shell. This approach is similar to the standard
description of the off-shell $W$-pair production in SAA. We have
taken into account the soft and hard initial state radiation and a
part of the virtual radiative corrections which are relevant in the
framework of the model.

It follows from our results that the model is applicable to
description of the near-threshold boson-pair production with LEP II
accuracy. We get the total cross-section which is in good accordance
with the experimental data; it coincides with the MC calculations
with a high precision. At the same time, the model provides a
compact analytical expression for the cross-section in terms of
convolution of the Born cross section with probability densities (or
mass distributions) of $W$ bosons. However, we did not fulfill the
detailed analysis of an accounting of the EW corrections, so this
rather phenomenological formalism can not be directly applied for
the precise description of the boson-pair production at high
energies and for future experiments at ILC. It is reasonable to
consider the possibility of improvement of the approach and its
applicability at the energies far from the near-threshold range. We
leave this analysis for a separate study.

\section*{Acknowledgments}

We would like to thank Vitaly Beylin and Gregory Vereshkov for
fluent discussions. Helpful correspondence with Stefan Dittmaier is
gratefully acknowledged. This work was supported in part by RFBR
Grants No. 07-02-91557 and No. 09-02-01149.

\end{document}